\documentclass[twocolumn,aps,floatfix,showpacs,superscriptaddress]{revtex4}

\usepackage{graphicx}
\usepackage{bm}
\begin{document}

\title{Sensitive measurement of forces at micron scale using Bloch oscillations of ultracold atoms}

\affiliation{Istituto Nazionale per Fisica per la Materia BEC-CRS
and Dipartimento di Fisica,\\ Universit\'a di Trento, I-38050
Povo, Italy}

\affiliation{\'Ecole Normale Sup\'erieure and Coll\`ege de France,
  Laboratoire Kastler Brossel, 24 Rue Lhomond 75231 Paris, France}

\affiliation{LENS and Dipartimento di Fisica, Universit\`a di Firenze,
  and INFM\\  Via Nello Carrara 1, 50019 Sesto Fiorentino, Italy }

\author{I. Carusotto}
\affiliation{Istituto Nazionale per Fisica per la Materia BEC-CRS
and Dipartimento di Fisica,\\ Universit\'a di Trento, I-38050
Povo, Italy}

\author{L. Pitaevskii}
\affiliation{Istituto Nazionale per Fisica per la Materia BEC-CRS
and Dipartimento di Fisica,\\ Universit\'a di Trento, I-38050
Povo, Italy}

\author{S. Stringari}

\affiliation{Istituto Nazionale per Fisica per la Materia BEC-CRS
and Dipartimento di Fisica,\\ Universit\'a di Trento, I-38050
Povo, Italy}

\affiliation{\'Ecole Normale Sup\'erieure and Coll\`ege de France,
  Laboratoire Kastler Brossel, 24 Rue Lhomond 75231 Paris, France}

\author{G. Modugno}

\affiliation{LENS and Dipartimento di Fisica, Universit\`a di Firenze,
  and INFM\\  Via Nello Carrara 1, 50019 Sesto Fiorentino, Italy }

\author{M. Inguscio}
\affiliation{LENS and Dipartimento di Fisica, Universit\`a di Firenze,
  and INFM\\  Via Nello Carrara 1, 50019 Sesto Fiorentino, Italy }

\begin{abstract}
We show that Bloch oscillations of ultracold fermionic atoms in
the periodic potential of an optical lattice can be used for a
sensitive measurement of forces at the micrometer length scale,
e.g. in the vicinity of dielectric surface. In particular, the
proposed approach allows to perform a local and direct measurement
of the Casimir-Polder force which is, for realistic experimental
parameters, as large as $10^{-4}$ gravity.
\end{abstract}
\pacs{
39.20.+q, 
03.75.Ss, 
03.75.Lm  
}

\date{\today}
\maketitle

The study of forces at small length scales constitutes one of the
challenges of current frontier research in physics. The interest
spans from fundamental issues, like the search of possible
deviations from newtonian gravity, to the phenomenology of forces
close to surfaces. Accurate investigations have been performed
using a variety of methods, as for instance the techniques based
on micro-cantilevers \cite{carugno,chiaverini}. The laser
production and manipulation of micron-size ultracold atomic
samples has provided a new tool for the investigation of forces at
small length scales \cite{hinds,shimizu}. More recently, also
trapped atomic Bose-Einstein condensates have been applied to the
study of forces close to surfaces \cite{vuletic,cornell,antezza}.

The possibilities offered by quantum degenerate gases have been
enriched by the combination with periodic trapping potentials, as
those produced by optical lattices. As early demonstrated
\cite{salomon,raizen,anderson}, this combination allows to study
the general phenomenon of Bloch oscillations \cite{bloch}. A key
feature of this phenomenon is that the measurement of an
oscillation frequency, which can be made very precise, can be
turned into a direct measurement of a force. An important step in
this direction has been the observation of temporally-resolved
Bloch oscillations of a degenerate Fermi gas of atoms trapped in
an optical lattice and subjected to gravity \cite{roati}. This has
shown the feasibility of a high sensitivity force sensor with
spatial resolution of the order of few microns.

In this paper we explore the effect of a weak inhomogeneous force
close to a surface, such as the Casimir-Polder force
\cite{casimir}, on atomic Bloch oscillations (BO). We make a
complete theoretical analysis for a Fermi gas trapped in an
optical lattice in proximity of a dielectric surface, using
realistic parameters from an existing experiment \cite{roati}. We
discover that this atomic Bloch oscillator realizes a powerful
sensor for the detection of forces at small length scales.

\begin{figure}[htbp]
\includegraphics[width=7cm,clip]{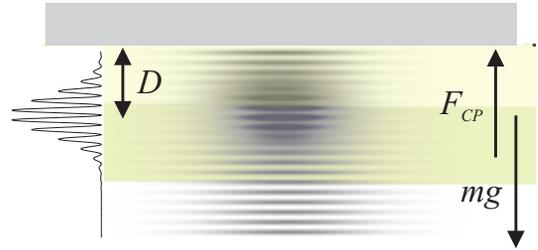}
\caption{Sketch of the physical situation considered in this
paper: an ultracold sample of fermionic atoms is trapped in a
vertical optical lattice in proximity of a dielectric surface and
performs Bloch oscillations under the combination of gravity and
of the Casimir-Polder force.} \label{fig1}
\end{figure}

The physical system that we consider here consists of a sample of
ultracold atoms trapped in a 1D lattice aligned along gravity, as
shown in Fig.\,\ref{fig1}. The atoms are initially cooled in a
harmonic potential, and then transferred to the optical lattice. A
harmonic horizontal confinement is assumed to be present, e.g.
given by the same red-detuned lattice beams. As soon as the
vertical harmonic confinement is switched off, the atoms start to
perform BO in the lattice potential under the action of gravity.
If now a surface is brought close to the atomic sample, additional
forces between the atoms and the surface will affect the dynamics
of BO. We will now investigate how such forces can be measured
from the shift of the Bloch oscillation frequency as the distance
between atoms and surface is varied.

In particular, we consider here the Casimir-Polder force, which is
exerted on an atom by the modified fluctuations of the
electromagnetic field in the vicinity of a dielectric
surface~\cite{casimir}. As a first step, we have worked out the
simplified model in which the Casimir-Polder (CP) force $F_{CP}$
is assumed to be spatially homogeneous. This is equivalent to
assuming that the size of the sample along the vertical direction
is small compared to the actual spatial variation of the force. In
this case the quasimomentum $q$ of the atoms simply evolves as
\begin{equation}
\hbar {\dot q}=F=mg+F_{CP} \label{eq:force}
\end{equation}
$F$ being the total force acting on the atom. As $q$ is defined
only modulo the Bragg momentum $q_B=2\pi/\lambda$ ($\lambda$ is
the wavelength of the optical lattice), each time $q$ gets to the
edge of the first Brillouin zone at $q_B$, it reappears on the
other side at $-q_B$. The period of BO is thus
$T_B=4\pi\hbar/F\lambda$. Following the evolution of BO for many
periods one can infer their frequency, and therefore the
Casimir-Polder force, with high accuracy. The fact that the
frequency shift here is proportional to the force and not to its
spatial gradient is an important difference of the present set-up
with respect to previous work \cite{antezza}, and opens new
possibilities for the measurement e.g. of radiation pressure forces
close to a heated surface~\cite{Henkel}. 
In \cite{antezza}
the effect of the CP force was investigated by looking at the
shifts of the collective oscillation frequencies of a
Bose-Einstein condensate  confined in a harmonic trap. In this
case the shift is proportional to the gradient of the force.
Experiments in this direction are being performed by the group of
E. Cornell at JILA \cite{cornellb}.

\begin{figure}[htbp]
\includegraphics[width=\columnwidth,clip]{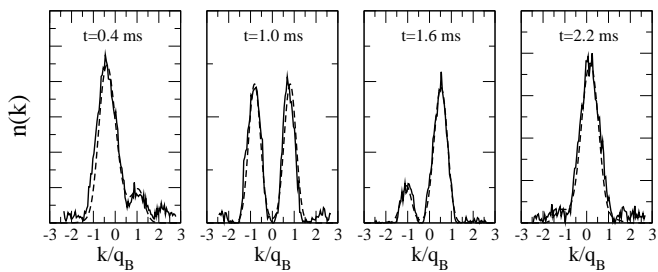}
\caption{Momentum distribution of a Fermi gas performing Bloch
  oscillations at different times within an oscillations period:
solid lines are experimental data, dashed lines are theoretical
  calculations including adiabatic switch-off of the lattice.
$E_F/2\delta=0.9$.}
\label{fig2}
\end{figure}

To observe BO one can study the momentum distribution of the
atomic sample after a ballistic expansion. The typical evolution
of the momentum is the one depicted in Fig.\,\ref{fig2}. At the
initial time $t=0$ the momentum distribution is centered at $k=0$,
with symmetric lateral peaks at multiples of $2q_B$ \cite{pedri}.
As time goes on, the comb of peaks moves rigidly with constant
velocity, while their relative intensity follows the Fourier
transform of a Wannier state~\cite{YuCardona} as an envelope. In
particular, when $q(t)$ reaches the edge of the first Brillouin
zone at $t=T_B/2$, the distribution contains a pair of symmetric
peaks at $\pm q_B$. At $t=T_B$, the distribution has regained the
initial shape. Note that the dynamics of BO in momentum space,
being governed by the evolution equation (\ref{eq:force}) for the
quasimomentum, is independent of the lattice height, while the
corresponding oscillations in real space have an amplitude of
order $\delta/F$, where $2\delta$ is the width of the first energy
band of the lattice. The presence of a strong known force, such as
gravity, is therefore necessary to reduce the amplitude of the
oscillation and consequently to achieve a high spatial resolution
in the measurement of a weak force. The use of a vertical geometry
is helpful also to prevent the atoms from hitting the surface
during the ballistic expansion.

This discussion, originally developed for a single particle, is
immediately extended to the case of a many-atom gas simply by
convolving the single particle prediction by the initial
quasi-momentum distribution of the gas: a fundamental requirement
to clearly follow BO is that the quasi-momentum spread of the
sample must not be much larger than $q_B$. This implies that the
atomic temperature must be at most of the order of $2\delta$. As
the system size is limited by the spatial resolution we intend to
obtain, a high density sample has to be used to increase the
number of atoms and thus improve the signal. This motivates the
use of a quantum degenerate gas.

Although a Bose-Einstein condensate would certainly offer a much
narrower initial quasi-momentum spread than a Fermi gas, it has
been shown in recent experiments~\cite{fallani} to suffer from
instabilities due to interactions, which quickly broaden the
quasi-momentum distribution and make the observation of BO
practically impossible in condensates at high density. As a matter
of fact, for the rest of the paper we will mainly consider a Fermi
gas: thanks to the absence of interactions this appears to be an
excellent candidate for this kind of experiments \cite{roati}. For
simplicity we will consider a Fermi gas at $T=0$, whose
quasi-momentum distribution is:
\begin{equation}
N_q=\frac{1}{2}\,\left(\frac{E_F-E_q}{\hbar\omega_r}\right)^2
\,\Theta(E_F-E_q)
\end{equation}
where $E_q$ is the energy of the Bloch state of quasi-momentum
$q$, $E_F$ is the Fermi energy, and $\omega_r$ is the frequency of
the radial harmonic trapping. As long as $E_F$ is smaller than
$2\delta$, the quasi-momentum distribution is localized in a
region around $q=0$ and is zero outside, while the contrast slowly
deteriorates for larger values of $E_F$. Notice that this behavior
is different from the case of a purely one-dimensional system,
where the contrast would be strictly zero as soon as $E_F \geq
2\delta$ \cite{pezze}.

The contrast can be improved by adiabatically switching off the
lattice potential at the end of the BO. In this case, the
population of the state of quasimomentum $q$ is projected onto the
first Brillouin zone without being weighted by the Fourier transform
of the Wannier state and at most two peaks are observed.
In Fig.\,\ref{fig2} we report the result of a theoretical analysis
taking into account the adiabatic switch-off of the lattice, which
reproduces with excellent accuracy the experimental data taken as
described in \cite{roati}.

\begin{figure}[htbp]
\includegraphics[width=6cm,clip]{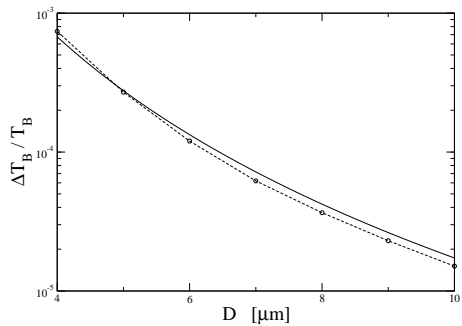}
\caption{Frequency shift of Bloch oscillations due to
Casimir-Polder corce. The solid line is the
prediction of Eq.\ref{eq:shift} based on the thermal potential
(\ref{eq:thermal}), circles are the result of a complete
numerical simulation as described in the text.
}
\label{fig3}
\end{figure}

Let us now study the effect of the CP force between the atoms and
the dielectric surface. A general discussion of its properties has
been recently given by~\cite{antezza}, and we refer to that paper
for the details of the calculations. In the simulations for the
present paper, the potential plotted in Fig.\,3 of~\cite{antezza}
has been used, which takes into
  account all the relevant regimes (van der Waals-London, Casimir-Polder,
  and thermal), in particular the thermal one
\begin{equation}
V_{CP}^{(th)}(z)=-\frac{k_B T \alpha_0}{4
z^3}\frac{\epsilon_0-1}{\epsilon_0+1}\,,
\label{eq:thermal}
\end{equation}
which dominates at distances larger than the thermal wavelength
$\hbar c/k_BT$ of the photon. Here $\alpha_0$ is the static atomic
polarizability, $\epsilon_0$ is the static dielectric constant of
the material composing the surface, which is placed at $z=0$, the
$z$ axis being oriented downwards. At the lowest level, one can
approximate the spatially varying force with a homogeneous one
equal in magnitude to the value of thermal force
(\ref{eq:thermal}) evaluated at the center of the cloud. In this
case, the BO period for an atomic cloud at a distance $D$ from the
surface is easily evaluated from eq.\ref{eq:force} and has a
relative shift
\begin{equation}
\frac{\Delta
T_B}{T_B}=-\frac{F_{CP}(D)}{mg}=\frac{0.1728}{D^4}(\rm \mu m)^4\,.
\label{eq:shift}
\end{equation}
The numerical value refers to the specific case of
potassium atoms and a sapphire surface with $\epsilon_0=9.4$ and
$T=300\,\textrm{K}$.
In Fig.\,\ref{fig3} we show the relative BO period shift
as a function of the distance $D$, computed according to the
approximated formula (\ref{eq:shift}).
Note how the effect of the Casimir potential ranges from $10^{-4}$ to
$10^{-5}$ at distances $D$ from 5 to 10 microns, which are
realistic for a possible experiment. This
situation is encouraging because such level of sensitivity has
been already demonstrated in a proof-of-principle experiment
\cite{roati}, and can be further improved by a few orders of
magnitude, as we will discuss later.

It is then worth performing a deeper analysis taking fully into
account the spatial inhomogeneity of the Casimir force over the
atomic cloud, its full $z$  dependence beyond the large distance
approximation (3) as well as the effect of a real-space motion of
the cloud during BO. We have indeed performed an exact numerical
calculation for the non-interacting many fermion problem, in the
combined potential of the trap, the gravity, the lattice and the
complete CP force as predicted in~\cite{antezza}. This has been
done by first obtaining the single-particle orbitals in the
initial potential of the trap, the lattice, and the CP force, by
then populating them according to a Fermi distribution, and by
finally propagating them via the Schr\"{o}dinger equation in the
combined potential of the lattice, the gravity and the CP force.
This procedure provides the density profile of the gas, as well as
its momentum and quasi-momentum distributions at all times. From
the center-of-mass of the quasi-momentum distribution, it is then
possible to extract a theoretical prediction for the BO frequency
shift.

\begin{figure}[htbp]
\includegraphics[width=6cm,clip]{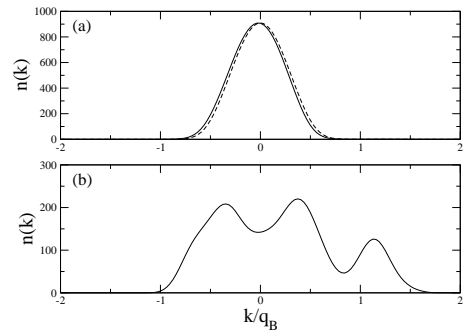}
\caption{Upper panel (a): momentum distribution of a Fermi gas at
a distance   $D=10\,\mu\textrm{m}$ from the surface at beginning
(dashed) and after 1000 periods   of Bloch   oscillations (solid).
Lower panel (b): momentum distribution of a BEC after 2 periods of
Bloch oscillations.} \label{fig4}
\end{figure}

The results of such a calculation are shown in Fig.\,\ref{fig3}
for realistic parameters: a situation has been considered in which
$2\cdot 10^4$~~$^{40}$K atoms are prepared in a lattice with
$\lambda=873\,\textrm{nm}$, with a vertical extension of about
$4\,\mu\textrm{m}$, i.e. $8$ lattice sites. To achieve a
favourable ratio $E_F/2\delta=0.9$ close to one we have chosen a
lattice height $s=5$, and a radial frequency of $10\,\textrm{Hz}$.
The vertical extension in the lattice is determined by the
confinement of the initial harmonic potential of a trap, which for
this specific case has a frequency of $400\,\textrm{Hz}$.
Justification for the slight difference with respect to the
analytical formula (\ref{eq:shift}) can be found in several
effects that have been neglected in the previous discussion, in
particular the spatial displacement (on the order of
$1\,\mu\textrm{m}$ in the direction of going further from the
surface) of the atomic cloud while performing BO, the difference
between the approximate thermal potential (\ref{eq:thermal}) and
the exact one of~\cite{antezza} and finally the spatial
inhomogeneity of the CP force across the atomic cloud. The
inhomogeneity of the force can have a further consequence: the
frequency shift of BO is in fact given by an average of the force
over the cloud profile, and a sort of inhomogeneous broadening is
to be expected. As this latter effect may affect the accuracy of
the measurement, in particular after the large number of
oscillations which are required for a high-precision measurement,
we have investigated the broadening of the momentum distribution
after a large number of BO periods. The results, reported in
Fig.\,\ref{fig4}a, show that the broadening is not yet significant
after 1000 periods.

For the sake of completeness, a complete analysis has been also
performed for the bosonic case by numerically solving the 1D
Gross-Pitaevskii equation with trap and atom number parameters
analogous to the Fermi case (that is with a peak density of
$n=4\cdot 10^{13}\,\textrm{cm}^{-3}$). The results are in
agreement with the experimental observations of~\cite{roati}: the
quasi-momentum distribution dramatically broadens as soon as the
quasi-momentum of the Bose condensate enters the dynamically
instable region~\cite{WuNiu} and after only a few BO periods, it
is not possible to follow the oscillations any longer
(Fig.\,\ref{fig4}b). To overcome this problem, one could think of
setting the scattering length $a$ to zero by means of a Feshbach
resonance~\cite{Feshbach}. As the characteristic time for the
onset of the instability is of the order of $\tau_{inst}\approx
\hbar/g\,n_{max}$~\cite{WuNiu}, $g\,n_{max}=4\pi\hbar^2\,a n_{max}
/m$ being the mean-field interaction energy, an upper bound on the
absolute value of $a$ is set by the fact that $\tau_{inst}$ has to
be much longer than the actual duration of the measurement. For
instance, a measurement lasting $100$ BO periods, performed with
the atomic density here considered, would require a reduction of
the scattering length of a factor of the order of $10^3$.

Let us now discuss the sensitivity that can be realistically
achieved in an experiment. In \cite{roati} we demonstrated an
overall sensitivity $\Delta T_B/T_B$=$10^{-4}$, by following about
100 BO periods of $^{40}$K atoms in a lattice with
$\lambda=873\,\textrm{nm}$. This sensitivity was limited by
various sources of decoherence due to the imperfect nature of the
optical lattice. Examples are the phase noise due to mechanical
vibrations of the retroreflecting mirror, or longitudinal
inhomogeneous forces due to the focusing of the lattice beams.
Most of these decoherence sources can be reduced in an optimized
experiment, thus extending the number of oscillations observable
with high contrast. The typical lifetime of the atomic sample of a
few seconds will however limit the realistic number of observable
BO periods to the order of 10$^3$. Since the sensitivity on the
measurement of a single BO period can be pushed to
10$^{-3}\div$10$^{-4}$, as typically achieved in this kind of
experiments \cite{biraben}, one can expect to obtain an overall
sensitivity $\Delta T_B/T_B$=$10^{-6}\div$10$^{-7}$.

This sensitivity would certainly allow to study a weak and
inhomogeneous force like the Casimir-Polder force already at
distances larger than 5\,$\mu$m from a dielectric surface. This
level of accuracy might also open to the investigation of possible
deviations from Newton's gravitational law at short distance
\cite{geraci}. In order to improve the probing of the force at
shorter distances one should reduce the width of the sample by
working, for example, with a stronger vertical confinement and
with a shorter lattice periodicity. The latter would indeed work
in favor of a higher resolution, without affecting the sensitivity
because, as already discussed, the measurable quantity ${\dot q}$
depends only on the applied force.

In conclusion we have discussed how Bloch oscillations can provide
a sensitive tool for the study of forces at the micrometer scale,
e.g. the Casimir-Polder force between atoms and dielectric
surfaces. The frequency shift of Bloch oscillations is
proportional to the applied force and at a distance of the order
of a few microns turns out to be well within the sensitivity of
realistic experiments.

We are grateful to M. Antezza for continuous stimulating
discussions on the Casimir-Polder forces. Discussions with G. Roati,
F. Ferlaino, G. Orso, C. Tozzo and F. Dalfovo are also
acknowledged. M.I. aknowledges the support of the Institute d'Optique,
Orsay, and the stimulating hospitality in the group af A.Aspect.
This work was supported by MIUR and by EU under
contract HPRICT1999-00111.

\end{document}